\documentstyle [12pt,a4,epsf] {article}

\begin{document}


\title {Kinetics of Surfactant Adsorption at Fluid/Fluid Interfaces:
        Non-Ionic Surfactants}
\author {Haim Diamant and David Andelman \\
         \\
         School of Physics and Astronomy   \\
         Raymond and Beverly Sackler Faculty of Exact Sciences \\
         Tel Aviv University, Ramat Aviv, Tel Aviv 69978, Israel \\
         \\}
\date{April 1996}
\maketitle


\vspace{1cm}
\begin{abstract}
\setlength {\baselineskip} {10pt}

We present a model treating the kinetics of adsorption of 
soluble surface-active molecules at the interface between an aqueous 
solution and another fluid phase.
The model accounts for both the diffusive transport inside the 
solution and the kinetics taking place at the interface using a 
free-energy formulation.
In addition, it offers a general method of calculating dynamic surface 
tensions.
Non-ionic surfactants are shown, in general, to undergo a 
diffusion-limited adsorption, in accord with experimental findings. 

\end{abstract}

\pagebreak

\setlength{\baselineskip}{10pt}
\setcounter{equation}{0}

Surface-active agents ({\em surfactants}) play a major role in
various fields, including petrochemical technologies, detergents,
biological systems, etc.
In some important cases, equilibrium properties of the 
surfactant adsorption are not sufficient, and knowledge of kinetic 
processes is required.
In particular, we mention processes of fast wetting, foaming
and the stability of liquid films.
The kinetics of surfactant adsorption have been addressed by numerous
experimental and theoretical studies, and various experimental 
techniques have been devised, primarily aimed at the measurement 
of dynamic interfacial tensions \cite{review}.

The pioneering theoretical work of Ward and Tordai \cite{WT} 
formulated a time-dependent relation between the surface density
of surfactants adsorbed at an interface and their concentration
at the sub-surface layer of solution, assuming a diffusive 
transport from the bulk solution.
Consequent theoretical works have focused on providing a second
closure relation between these two variables.
Various relations have been suggested, resembling equilibrium 
isotherms \cite{isotherm,Hansen}, or having a
kinetic differential form \cite{kinetic,Lin1}.
Such theories have been quite successful in describing the 
experimentally observed adsorption of common non-ionic surfactants.
Yet, they suffer from several drawbacks:
(i) The closure relation between the surface density and sub-surface
concentration, which expresses the kinetics taking place just at the 
interface, is introduced as an {\em external} boundary condition, 
and does not uniquely arise from the model itself;
(ii) The calculated dynamic surface tension relies on 
an {\em equilibrium} equation of state, and assumes that it also 
holds out of equilibrium \cite{Fordham};
(iii) Similar theories cannot be successfully extended to describe more
complicated, {\em ionic} surfactant solutions
\cite{ionic}.
In the current paper we would like to briefly present an alternative
approach, overcoming these three drawbacks.

Consider an interface between an aqueous solution of non-ionic 
surfactants and an air or oil phase at $x=0$.
At $x \rightarrow \infty$, the solution is in contact with a bulk
reservoir of surfactants, where the chemical
potential and surfactant volume fraction are fixed to be 
$\mu_b$ and $\phi_b$, respectively.
We consider a dilute solution, {\it i.e.} the surfactant volume
fraction is much smaller than unity throughout the solution.
The concentration is also smaller than the critical
micelle concentration ({\em cmc}), so the surfactants are dissolved 
only as monomers.
At the interface itself, however, the volume fraction may become
large. 

We write the excess in free energy per unit area due to the 
interface ({\it i.e.} the change in interfacial tension) as a 
functional of the surfactant volume fraction in the bulk solution, 
$\phi(x>0)$, and its value at the interface, $\phi_0$,
$
  \Delta \gamma [\phi] = \int_0^\infty \Delta f[\phi(x)] dx + 
        f_0 (\phi_0)
$.
The first term is the contribution from the bulk solution,
whereas the second is the contribution from the interface itself.
The sharp, ``step-like" profile considered has led us to treat
the bulk solution and the interface as two coupled sub-systems, 
rather than a single one \cite{Tsonop}.
The bulk sub-system is considered as an ideal, dilute solution,
including only ideal entropy of mixing and the contact with 
the reservoir, and neglecting gradient terms,
\begin{equation}
  \Delta f(\phi) = \{ T [ \phi\ln\phi - \phi - 
    (\phi_b\ln\phi_b - \phi_b) ] - \mu_b 
    (\phi - \phi_b) \} / a^3,
 \label{Df}
\end{equation}
where $a$ denotes the surfactant molecular dimension and $T$ the 
temperature (we set the Boltzmann constant to unity).
However, since $\phi_0$ may become much larger than $\phi(x>0)$,
we must take into account at the interface the finite molecular size and
interactions between surfactant molecules,
\begin{equation}
  f_0(\phi_0) = \{ T [ \phi_0\ln\phi_0 + (1-\phi_0)
    \ln(1-\phi_0) ] - \alpha\phi_0 - (\beta/2)\phi_0^2 - 
    \mu_1\phi_0 \} / a^2.
 \label{f0}
\end{equation}
The first term is due to the entropy of mixing, recalling that $\phi_0$
is not necessarily small;
the second accounts for the energetic preference of the surfactants to 
lie at the interface;
the third is the energy of lateral attraction between neighboring 
surfactants at the interface;
and the last term accounts for the contact with the solution adjacent 
to the interface, where the chemical potential is $\mu_1$.
 
Variation of $\Delta\gamma$ with respect to $\phi(x)$ yields the excess 
in chemical potential at a distance $x$ from the interface,
$
  \Delta \mu(x) = a^2 \delta\Delta\gamma / \delta\phi(x) =
                 T \ln\phi(x) - \mu_b 
$ for $x>0$, and
$  \Delta \mu_0 = a^2 \delta\Delta\gamma / \delta\phi_0 =
                 T \ln [\phi_0/(1-\phi_0)] - \alpha 
                 - \beta\phi_0 - \mu_1
$.
From these equations we can deduce, as expected,
$\mu_b=T\ln\phi_b$ and $\mu_1=T\ln\phi_1$,
where $\phi_1$ denotes the sub-surface volume fraction.
 
\noindent
{\it Thermodynamic Equilibrium}.~~~ In equilibrium the chemical potential 
is equal to $\mu_b$ throughout the entire system (the variations of 
$\Delta\gamma$ vanish).
From the variation with respect to $\phi(x)$ 
we obtain the equilibrium profile, $\phi(x)\equiv\phi_b$ for $x>0$,
and from the variation with respect to $\phi_0$ 
the equilibrium adsorption isotherm,
\begin{equation}
  \phi_0 = \phi_b / [\phi_b + {\rm e}^{-(\alpha+\beta\phi_0)/T}].
 \label{Frumkin}
\end{equation}
We have recovered the {\em Frumkin adsorption isotherm}, which
reduces to the well-known {\em Langmuir adsorption isotherm} 
\cite{Adamson} when the interaction term is neglected ($\beta=0$).
From Eq.~(\ref{f0}) and the above variations one also
obtains the equilibrium equation of state,
$
  \Delta\gamma = [ T\ln(1-\phi_0) + (\beta/2) \phi_0^2 ] / a^2
$,
expressing the equilibrium dependence of the surface pressure, 
$\Pi=-\Delta\gamma$, on the surface coverage, $\phi_0$, 
according to a lattice-gas model.

\noindent
{\it Out of Equilibrium}.~~~ We assume proportionality between 
velocities and the chemical potential gradient \cite{Langer}, 
and take the surfactant mobility to be $D/T$, according to
the Einstein relation ($D$ being the surfactant diffusivity).
At positions not adjacent to the interface
we obtain for the surfactant current density
$j(x)=-\phi(D/T)\partial\mu/\partial x=-D\partial\phi/\partial x$,
and applying the continuity condition, 
$\partial\phi/\partial t = -\partial j/\partial x$, 
get the ordinary {\em diffusion equation},
$
  \partial\phi/\partial t = D \partial^2\phi/\partial x^2
$.

The proximity of the interface requires a more careful treatment.
First, we discretize expression for $\Delta\gamma$ on a lattice with cells 
of length $a$, 
$\Delta\gamma[\phi]=a\sum_{i=1}^\infty\Delta f(\phi_i)+f_0(\phi_0)$,
where $\phi_i\equiv\phi(x=ia)$. Discretized current densities, $j_i$,
can be similarly defined.
Since we do not allow molecules to leave the interface towards
the other phase ($j_0=0$), we have from continuity
$\partial\phi_0/\partial t = - j_1/a$,
and can therefore write
$
  \partial\phi_1/\partial t = -(j_2-j_1)/a =
        (D/a) \partial\phi/\partial x|_{x=a}
        - \partial\phi_0/\partial t
$.
Applying the Laplace transform to the equations for
$\partial\phi/\partial t$ and $\partial\phi_1/\partial t$
while assuming an initial uniform state, 
$\phi(x,t=0) \equiv \phi_b$, a relation is obtained between the 
surface coverage and sub-surface volume fraction,
\begin{equation}
  \phi_0(t) = (1/a) \sqrt{D/\pi} [ 2\phi_b\sqrt{t} - 
          \int_0^t \phi_1(\tau) (t-\tau)^{-1/2} d\tau ]
          + 2\phi_b - \phi_1
 \label{WT}
\end{equation}
This relation is similar to the classical result of Ward and Tordai 
\cite{WT}, except for the term $2\phi_b - \phi_1$. 
The difference is due to fine details we have considered near
the interface and our initial condition, and it vanishes when $a$ 
goes to zero.
Finally, we find the equation determining the kinetics at the interface 
itself,
\begin{equation}
  \partial\phi_0/\partial t = \phi_1 D (\mu_1-\mu_0) / a^2 T 
    = (D/a^2) \phi_1 \{ \ln[\phi_1(1-\phi_0)/\phi_0] + \alpha/T
    + \beta\phi_0/T \}
 \label{dp0dt}
\end{equation}
Note, that the conditions at the interface are very different from those
inside the aqueous solution, and the diffusivities, $D$, appearing in 
the equations above, cannot be expected to have 
strictly the same value. 
Solution of Eqs. (\ref{WT}) and (\ref{dp0dt}) allows one to find the
time-dependent surface coverage, $\phi_0(t)$.

By writing the above equations, we have separated the kinetics of
the system into two coupled kinetic processes. 
The first takes place inside the bulk solution and is described by 
Eq.~(\ref{WT}),
whereas the second takes place at the interface and is described by 
Eq.~(\ref{dp0dt}).
Two limiting cases correspond to the relative speed of these 
two processes.
(i) {\em Diffusion-limited adsorption} applies when the process 
  inside the solution is much slower than the one at the interface.
  One can then assume that the interface is in constant equilibrium
  with the adjacent solution, {\it i.e.} the variation with respect to $\phi_0$
  vanishes, so $\phi_0$ immediately responds to changes in $\phi_1$ via 
  the equilibrium isotherm.
(ii) {\em Kinetically limited adsorption} takes place when the 
  kinetic process at the interface is the slower one.
  In this case, the solution is assumed to be in constant 
  equilibrium with the bulk reservoir, {\it i.e.} the variation 
  with respect to $\phi(x)$ vanishes [$\phi(x>0)=\phi_b$], 
  and $\phi_0$ changes with time   according to Eq.~(\ref{dp0dt}).

Looking at Eq.~(\ref{WT}) we can identify the 
{\em time scale of diffusion} for attaining the equilibrium coverage, 
$\phi_{0,eq}$,
\begin{equation}
  \tau_d = (\phi_{0,eq}/\phi_b)^2 a^2/D
 \label{td}
\end{equation}
Characteristic values of $a^2/D$ correspond to very short times (on
the order of nanoseconds), but the prefactor of 
$(\phi_{0,eq}/\phi_b)^2$ is typically very large (on the 
order of, say, $10^{11}$). 
Thus, the diffusion time scales may reach minutes, as indeed observed
in practice. 
In order to estimate the {\em time scale of kinetic adsorption} at the
interface, we examine the asymptotic behavior of Eq~(\ref{dp0dt}) 
close to equilibrium and find
\footnote
{
Close to equilibrium we can also write Eq.~(\ref{dp0dt}) as
$\partial\phi_0/\partial t \simeq (D\phi_b/a^2\phi_{0,eq}) [
   {\rm e}^{(\alpha+\beta\phi_0)/T}\phi_1(1-\phi_0) - \phi_0 ]$,
which coincides with the adsorption-desorption form of the Frumkin (or 
Langmuir, when $\beta=0$) kinetic equation used by previous authors
\cite{kinetic,Lin1}.
}
 $\phi_{0,eq}-\phi_0(t) \sim {\rm e}^{-t/\tau_k}$, with
\begin{equation}
  \tau_k \simeq (\phi_{0,eq}/\phi_b)^2 (a^2/D) 
     {\rm e}^{-(\alpha+\beta\phi_{0,eq})/T}
 \label{tk}
\end{equation}
Since the value of $D$ at the interface is not expected to be 
drastically smaller than that inside the solution,      
comparison of Eqs.~(\ref{td}) and (\ref{tk}) leads to the conclusion
that $\tau_d>\tau_k$.
This result is somewhat expected, since we did not include any
potential barrier in $f_0$. Adding such a barrier, $+\epsilon\phi_0$,
to Eq.~(\ref{f0}) will result in a factor of ${\rm e}^{\epsilon/T}$ in
$\tau_k$, describing a kinetic limitation.
We thus expect, in general, that {\em non-ionic surfactants 
should exhibit diffusion-limited adsorption}.
This, indeed, has been observed for quite a large number of non-ionic
surfactants \cite{review}.
\footnote
{
Note, that in the discussion above we have completely neglected a 
third time scale --- that needed for lateral diffusion and molecular 
re-orientation at the interface. 
If, however, due to certain molecular constraints, this time scale
is no longer negligible, exceptions to the above conclusions are to 
be expected \cite{lateral}.
}
The ``footprint" of diffusion-limited adsorption is the asymptotic
time dependence \cite{Hansen}
$\phi_{0,eq}-\phi_0(t) \sim  \sqrt{\tau_d/t}$.
Any dependence between the surface tension and surface coverage 
will lead to a similar asymptotic time dependence of the dynamic 
surface tension as well. 
Four examples of experimental results are given in Fig.~1, all
exhibiting the expected asymptotic $t^{-1/2}$ behavior.
\begin{figure}[tbh] 
\epsfysize=15\baselineskip
\centerline{\hbox{\epsffile{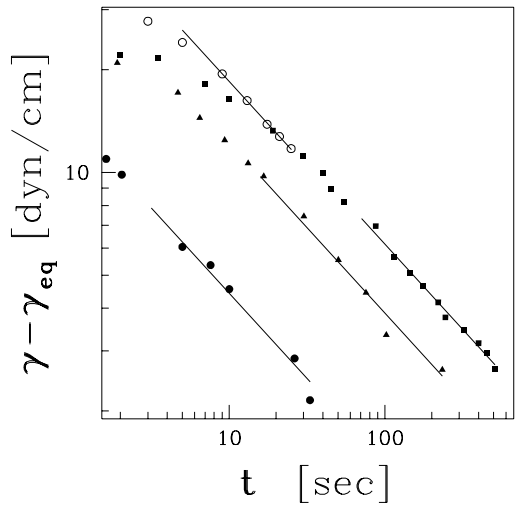}} }
\caption[]
{\footnotesize
  A variety of non-ionic surfactants were experimentally found to 
   exhibit diffusion-limited adsorption. 
   Four examples of dynamic surface tension measurements are given:
   $9.49\times 10^{-5}$M of decyl alcohol (open circles),
   from Ref.~\cite{AddHutch};
   $2.32\times 10^{-5}$M of Triton X-100 (squares),
   from Ref.~\cite{Lin1};
   $6\times 10^{-5}$M of C$_{12}$EO$_8$ (triangles)
   and $4.35\times 10^{-4}$M of C$_{10}$PY (solid circles),
   both from Ref.~\cite{Hua}.
   Note the asymptotic $t^{-1/2}$ behavior, characteristic
   of a diffusion-limited adsorption, and shown by the solid fitting 
   lines.
}
\end{figure} 

We return now to the interfacial tension during the process of
diffusion-limited adsorption.
As stated above, in this limit the interfacial contribution, 
$f_0(\phi_0)$, is at its minimum all the time, and $\Delta\gamma$ is
given therefore by
$\int_0^\infty \Delta f[\phi(x)] dx+ 
[T\ln(1-\phi_0)+(\beta/2)\phi_0^2]/a^2$.
If, in addition, we neglect the bulk contribution (recalling that it
completely vanishes at equilibrium)
we are left with the equilibrium equation of state.
Therefore, {\em relating the surface tension to surface coverage
via the equilibrium equation of state approximately holds also out of 
equilibrium}.
Note, that this statement is valid only in the case of 
{\em diffusion-limited adsorption}, where $f_0$ is at its minimum 
during the whole process.
The dependence of $\Delta\gamma$ on $\phi_0$, as defined by 
the equation of state, is shown in Fig.~2a. 
\begin{figure}[tbh] 
\epsfysize=15\baselineskip
\centerline{\hbox{\epsffile{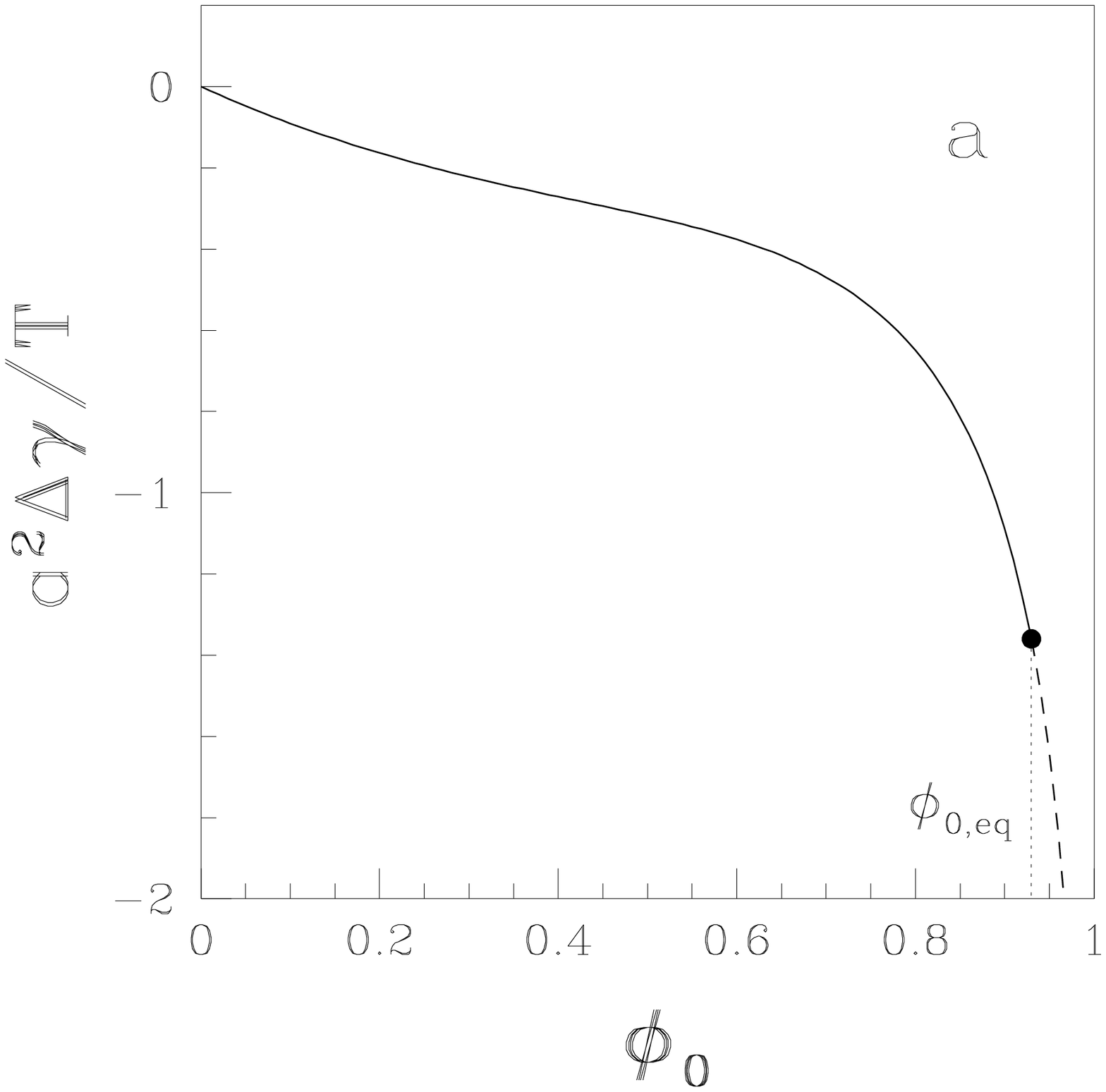}} 
                   \epsfysize=15\baselineskip \hbox{\epsffile{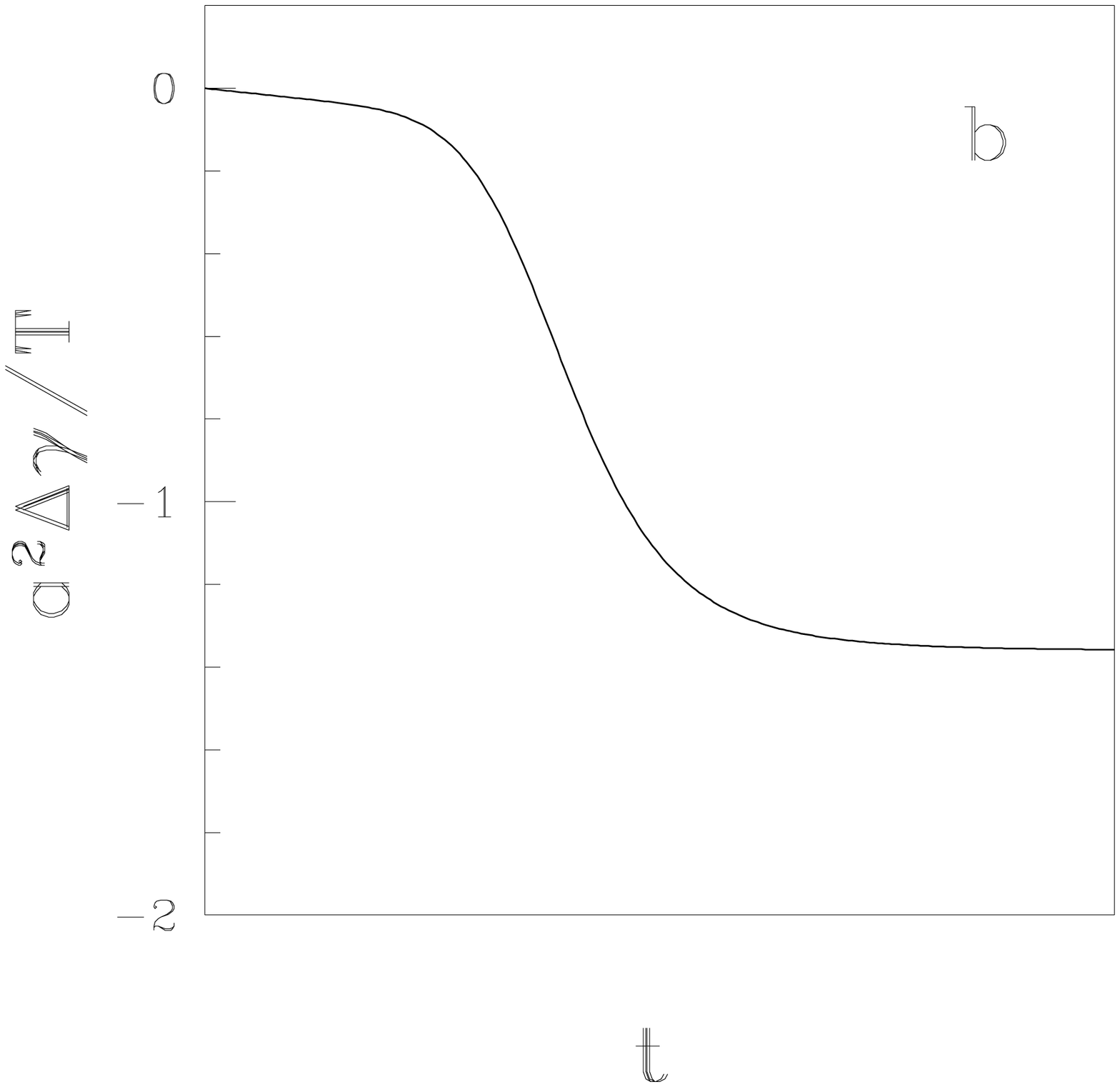}}}
\caption[]
{\footnotesize  
  (a) The dependence between surface tension and surface coverage in
       a diffusion-limited adsorption. The energy constants are set to the
       (realistic) values $\alpha=12T$ and $\beta=3T$;
   (b) Schematic time dependence of the surface tension, as expected
       from a dependence $\Delta\gamma(\phi_0)$ such as in (a).
}
\end{figure} 
Since $\phi_0$ constantly increases with time, we expect the {\em time} 
dependence of $\Delta\gamma$ to qualitatively resemble the curve 
depicted in Fig.~2b.
This, indeed, resembles the curves found in experiments 
({\it e.g.} \cite{Lin1}).
Note the almost constant slope in the beginning of the process;
the surface coverage significantly changes without a corresponding 
change in the surface tension. 
It is a result of the competition between the entropy and interaction 
terms appearing in the equation of state.
The surface tension will start falling roughly when the second
derivative of $\Delta\gamma$ with respect to $\phi_0$ changes sign
(see Fig.~2a), {\it i.e.} when
$1-\phi_0 \sim (\beta/T)^{-1/2}$.
As one examines surfactant solutions of increasing bulk 
concentrations, this surface coverage will be attained earlier along
the process, and the initial plateau will shrink, 
until finally vanishing behind the finite experimental resolution.
This trend is indeed observed experimentally \cite{Lin2}.

We have presented above an alternative model for the kinetics of 
non-ionic surfactant adsorption at fluid/fluid interfaces.
It is a more complete model, in the sense that the 
kinetics in the {\em entire} system, in the bulk solution as well as
at the interface, are described without the need for an additional, 
externally inserted boundary condition.
We have shown that relating the dynamic surface tension to surface coverage 
via the equilibrium equation of state, a procedure employed by 
practically all previous authors, is justified only in the case of 
diffusion-limited adsorption. 
Since the adsorption of non-ionic surfactants is generally
limited by diffusion, this assumption did not affect the validity
of previous theories.
We do not expect similar theories to be applicable to
{\em kinetically limited} systems, such as salt-free ionic 
surfactant solutions \cite{ionic}.
In contrast, our model allows, using the definition of $\Delta\gamma$
given above, for the calculation of dynamic surface tensions 
regardless of such limitations.
Like any other free-energy formulation, the model can be easily 
extended to include additional interactions.
A natural candidate should be the electrostatic interaction, {\it i.e.}
extension of the model to {\em ionic} surfactants.
This problem will be addressed in a future paper \cite{next}.

\vspace{1cm}
\newlength{\tmp}
\setlength{\tmp}{\parindent}
\setlength{\parindent}{0pt}
{\em Acknowledgments}
\setlength{\parindent}{\tmp}

We are indebted to D.~Langevin and A.~Bonfillon-Colin for introducing
us to the problem of dynamic surface tension, sharing with us their 
unpublished experimental data and for illuminating discussions.
We also benefited from discussions with N.~Agmon.
Support from the German-Israeli Foundation (G.I.F.)
under grant No.~I-0197 and the US-Israel Binational Foundation (B.S.F.)
under grant No.~94-00291 is gratefully acknowledged.


\end{document}